\documentclass[preprint,showpacs,preprintnumbers,amsmath,amssymb]{revtex4}
\usepackage[utf8]{inputenc}
\usepackage[dvips]{graphicx}
\usepackage{rotating}
\usepackage{hyperref}
\usepackage{subfig}
\usepackage[normalem]{ulem}
\usepackage{color}
\definecolor{orange}{rgb}{1,0.5,0}

\definecolor{red}{rgb}{1,0,0}
\usepackage[T1]{fontenc}
\usepackage{amsmath}
\usepackage{slashed}

\newcommand{\tvect}[2]{%
\ensuremath{\Bigl(\negthinspace\begin{smallmatrix}#1\\#2\end{smallmatrix}\Bigr)}}

\begin{document}

\begin{center}
{\large \bf Photons coming from an opaque obstacle as a manifestation of heavy neutrino decays}\\ \vspace{.9cm}
\end{center}
\author{Mat\'ias M. Reynoso, Ismael Romero and Oscar A. Sampayo}
\affiliation{Instituto de Investigaciones F\'{\i}sica de Mar del Plata (IFIMAR)\\
CONICET, UNMDP\\ Departamento de F\'{\i}sica,
Universidad Nacional de Mar del Plata \\
Funes 3350, (7600) Mar del Plata, Argentina}

\begin{abstract}
{\small Within the framework of physics beyond the standard model we study the possibility that mesons produced in the atmosphere by the cosmic ray flux, decay to heavy Majorana neutrino and these mostly to  photons in the low mass region. We study the photon flux produced by sterile Majorana neutrinos ($N$) decaying after passing through a massive and opaque object such as a mountain. 
In order to model the production of $N$'s in the atmosphere and their decay to photons, we consider the interaction between the Majorana neutrinos and the standard matter as modeled by an effective theory. We then calculate the heavy neutrino flux originated by the decay of mesons in the atmosphere.
The surviving photon flux, originated by $N$ decays, is calculated using transport equations that include the effects of Majorana neutrino production and decay. }
\end{abstract}

\maketitle

\section{\bf Introduction}

The neutrino sector has provided through the discovery of neutrino flavor oscillations the most compelling evidence for physics beyond the Standard Model ($SM$). 
However, other mysteries related with the same sector are still open. In particular, the tiny ordinary neutrino masses problem, for which the seesaw mechanism stays as one of the most straightforward ideas for solving it \cite{Minkowski:1977sc, Mohapatra:1979ia, Yanagida:1980xy, GellMann:1980vs, Schechter:1980gr, Kayser:1989iu}. This mechanism introduces right handed sterile neutrinos that, as they do not have distinct particle and antiparticle degrees of freedom, can have a Majorana mass term leading to the tiny known masses for the standard neutrinos, as long as the Yukawa couplings between the right handed Majorana neutrinos and the standard ones remain small. 
Even for the low masses range for $N$ here considered, the simplest Type-I seesaw scenario leads to a negligible left-right neutrino mixing $\vert U_{lN}\vert^2 \sim m_{\nu}/M_N \sim 10^{-9}$ \cite{deGouvea:2015euy, Deppisch:2015qwa, delAguila:2008ir}. { The mixing $U_{lN}$ weighs the couplings of $N$ with the S.M. particles and in particular with charged leptons through the $V-A$ interaction 
\begin{equation}\label{wlnu}
\mathcal{L}=-\frac{q}{\sqrt{2}} U_{lN} \bar N^c \gamma{\mu} P_L l W^+_{\mu} \; + \; hc
\end{equation}
}
Thus, as suggested in \cite{delAguila:2008ir}, the detection of Majorana neutrinos ($N$) would be a signal of physics beyond the minimal seesaw mechanism, and its interactions could be better described in a model independent approach based on an effective theory. We consider a simplified scenario with only one Majorana neutrino $N$ and negligible mixing with the $\nu_{L}$. 
In addition, the effective operators here presented allow the $N$-decay to one neutrino plus one photon. This decay channel  could account, in a particular
parameter region, for some neutrino related problems as the 
MiniBOONE and \cite{AguilarArevalo:2007it, AguilarArevalo:2008rc} and SHALON \cite{Sinitsyna:2013hmn} anomaly.
The Majorana neutrino effective phenomenology regarding the relevant $N$ decay modes and interactions is treated in \cite{Duarte:2015iba, Duarte:2016miz}.

In the present work we study the possibility that mesons produced in the atmosphere by the cosmic ray, decay to heavy Majorana neutrino and these mostly to  photons in the low mass region. We consider the scenario in which this radiative decay could be detected as a photon flux coming from an opaque obstacle such as mountain, which will stop the photons produced before the obstacle as well as  any other photons originated by another mechanism, leaving an observable survival photon flux generated after the obstacle.

In Sec. \ref{modelo}, we briefly describe the effective operator approach. In Sec.\ref{photonproduction}, we present the production mechanism of heavy neutrinos ($N$) by meson decay and the $N$ decay to photons. In Sec.\ref{resultados},
we discuss the bounds on the effective operators coming from different experiment as $0\nu\beta\beta$ decay, colliders results, meson decay, Super-K and astrophysical observations,  in the mass range of tens of MeV. We calculate the number of events of photons to be observed by a Cherenkov telescope at different distances from the opaque obstacle. We show the results as a contour plot for the number of events and include the region allowed by the bounds.
We leave to the Appendix \ref{ap1} the study of meson decay to $N$ in the effective formalism and in the Appendix \ref{ap2} we present the calculation for the $N \rightarrow \nu \gamma$ decay in the LAB frame.
 Finally, in Sec.\ref{conclusiones}, we present our conclusions.

\section{\bf Effective Majorana interactions \label{modelo}}

In this work, we study the observable effects of a heavy sterile Majorana neutrino $N$ decaying to photons after passing through a massive and opaque object such as a mountain. Thus, we 
need to model the interactions of $N$ with ordinary matter in order to describe the $N$ production by meson decay in the atmosphere and the subsequent $N$-decay to photons.
Being $N$ a $SM$  singlet, its only possible re-normalizable interactions with $SM$ fields involve Yukawa couplings. But, as we discussed in the introduction, these couplings must be very small in order to accommodate the observed tiny ordinary $\nu$ masses. In this work, we take an alternative approach, considering that the sterile $N$ interacts with the light neutrinos by higher dimension effective operators, and take this interaction to be dominant in comparison with the mixing through the Yukawa couplings. In this sense, we depart from the usual viewpoint in which the sterile neutrinos mixing with the standard neutrinos are assumed to govern the $N$ production and decay mechanisms \cite{Atre:2009rg, delAguila:2007qnc}.   

We parameterize the effects of new physics by a set of effective operators $\mathcal{O}$ constructed with the standard model and the Majorana neutrino fields, satisfying the $SU(2)_L \otimes U(1)_Y$ gauge symmetry \cite{delAguila:2008ir}.  The effect of these operators is suppressed by inverse powers of the new physics scale $\Lambda$, which is not necessarily related to the Majorana neutrino mass $m_{N}$. The total Lagrangian is organized as follows:
\begin{eqnarray}
\mathcal{L}=\mathcal{L}_{SM}+\sum_{n=6}^{\infty}\frac1{\Lambda^{n-4}}\sum_{\mathcal{J}} \alpha^{(i)}_{\mathcal{J}} \mathcal{O}_{\mathcal{J}}^{(n),i}
\end{eqnarray}
where $\mathcal{J}$ is the label of the operator, $n$ their dimension and $i$ the family.

For the considered operators, we follow \cite{delAguila:2008ir} starting with a rather general effective Lagrangian density for the interaction of right handed Majorana neutrinos $N$ with bosons, leptons and quarks. We list the dimension $6$ operators that can be generated at tree level or one-loop level in the unknown fundamental ultraviolet theory, and which are baryon-number conserving.
The first subset includes operators with scalar and vector bosons (SVB),
{
\begin{eqnarray} \label{eq:ope1}
\mathcal{O}^{(6),i}_{LN\phi}=(\phi^{\dag}\phi)(\bar L_i N \tilde{\phi}), \;\; \mathcal{O}^{(6),i}_{NN\phi}=i(\phi^{\dag}D_{\mu}\phi)(\bar N
\gamma^{\mu} N), \;\; \mathcal{O}^{(6),i}_{Ne\phi}=i(\phi^T \epsilon D_{\mu} \phi)(\bar N \gamma^{\mu} l_i)
\end{eqnarray}
}
and a second subset includes the baryon-number conserving 4-fermion contact terms:
{
\begin{eqnarray} \label{eq:ope2}
\mathcal{O}^{(6),i}_{duNe}&=&(\bar d_i \gamma^{\mu} u_i)(\bar N \gamma_{\mu} l_i) , \;\; \mathcal{O}^{(6),i}_{fNN}=(\bar f_i \gamma^{\mu}
f_i)(\bar N \gamma_{\mu}
N), \;\; \mathcal{O}^{(6),i}_{LNLe}=(\bar L_i N)\epsilon (\bar L_i l_i),
\nonumber \\
\mathcal{O}^{(6),i}_{LNQd}&=&(\bar L_i N) \epsilon (\bar Q_i
d_i), \;\; \mathcal{O}^{(6),i}_{QuNL}=(\bar Q_i u_i)(\bar N L_i) , \;\; \mathcal{O}^{(6),i}_{QNLd}=(\bar Q_i N)\epsilon (\bar L_i d_i), 
\nonumber \\
\mathcal{O}^{(6),i}_{LN}&=&|\bar N L_i|^2 , \;\; \mathcal{O}^{(6),i}_{QN}=|\bar Q_i N|^2
\end{eqnarray}
where $l_i$, $u_i$, $d_i$ and $L_i$, $Q_i$ denote, the right handed $SU(2)$ singlets and the
left-handed $SU(2)$ doublets, respectively for the family $i$.}
The following one-loop level generated operators coefficients are naturally suppressed by a factor $1/16\pi^2$ \cite{delAguila:2008ir, Arzt:1994gp}:
{
\begin{eqnarray}
\mathcal{O}^{(5),i}_{NNB} & = & \bar N \sigma^{\mu\nu} N^c B_{\mu\nu}, \cr \mathcal{O}^{(6),i}_{ N B} = (\bar L_i \sigma^{\mu\nu} N) \tilde
\phi B_{\mu\nu} , && \mathcal{O}^{(6),i}_{ N W } = (\bar L_i \sigma^{\mu\nu} \tau^I N) \tilde \phi W_{\mu\nu}^I , \cr \mathcal{O}^{(6),i}_{ D N} =
(\bar L_i D_\mu N) D^\mu \tilde \phi, && \mathcal{O}^{(6),i}_{ \bar D N} = (D_\mu \bar L_i N) D^\mu \tilde \phi \ . \label{eq:ope3}
\end{eqnarray}
}

As we will show these operators contribute to $N$ production by meson decay in the atmosphere and the subsequent $N$ decay to photons.

\begin{figure}
\begin{center}
\includegraphics[width=0.8\textwidth]{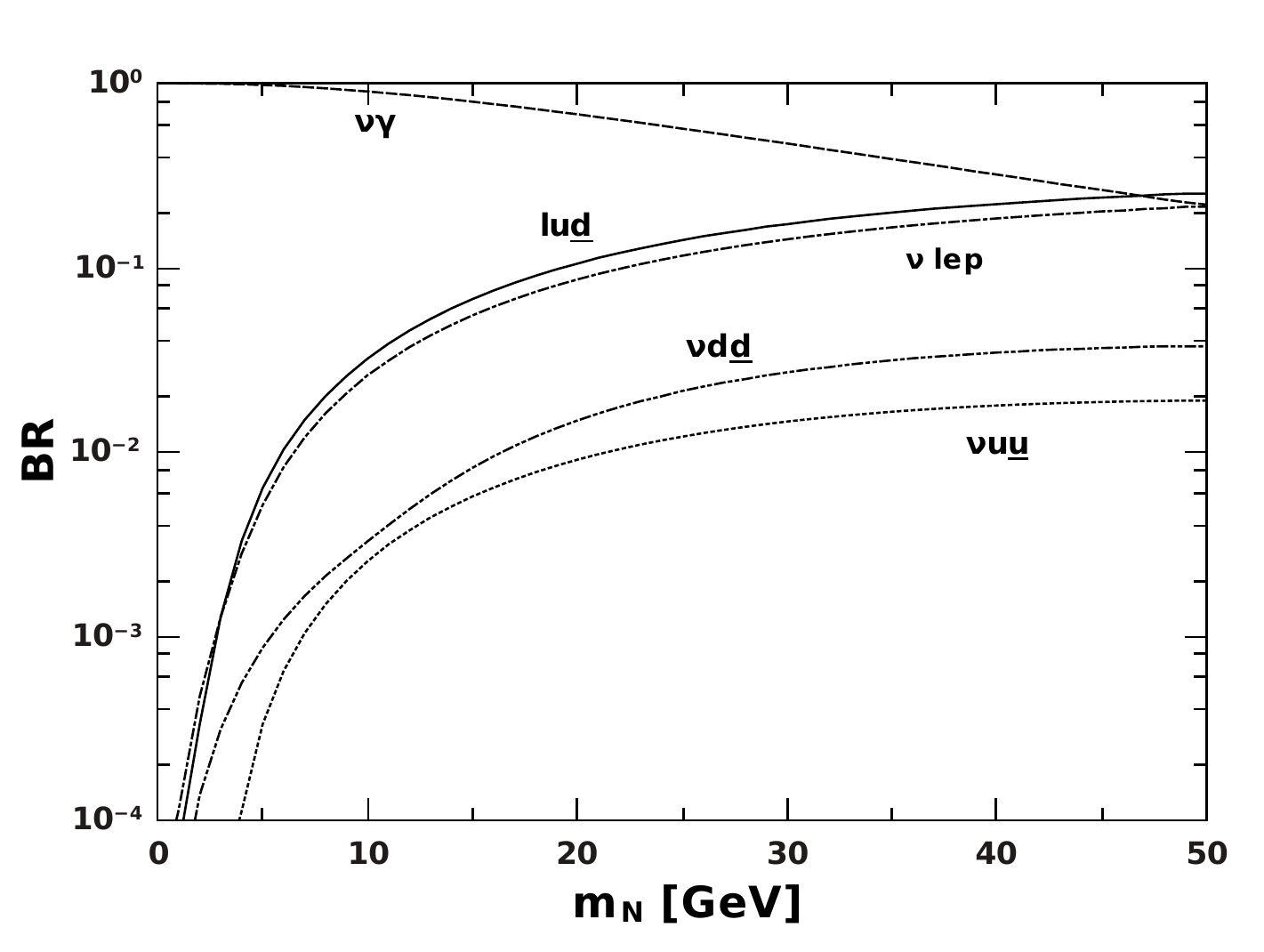}
\hspace{4cm}
 \caption{\label{fig:branching} Branching ratios for the main decay channels and for the couplings discussed in the text.} 
\end{center}
\end{figure}

In order to obtain the necessary interactions, we derive the relevant pieces of the effective Lagrangian terms involved in the calculations. We take the scalar doublet after spontaneous symmetry breaking as $\phi=\tvect{0}{\frac{v+h}{\sqrt{2}}}$. We have contributions to the effective Lagrangian coming from \eqref{eq:ope1}, related to the spontaneous symmetry breaking process:
\begin{eqnarray}\label{leff_svb}
 \mathcal{L}^{tree}_{SVB} &=& \frac{1}{\Lambda^2}\left\{
\alpha_Z (\bar N_R \gamma^{\mu} N_R) \left( \frac{v m_Z}{2} Z_{\mu} \right)  \right.
 - \left. \alpha^{(i)}_W (\bar N_R \gamma^{\mu} l_{R,i})\left(\frac{v m_{W}}{\sqrt{2}}W^{+}_{\mu} \right) + \cdots + h.c. \right\}, 
\end{eqnarray}
and the four-fermion interactions involving quarks and leptons from (\ref{eq:ope2})
{
\begin{eqnarray}\label{leff_4-f}
\mathcal{L}^{tree}_{4-f}&=& \frac{1}{\Lambda^2} \left\{ \alpha^{(i)}_{V_0} \bar d_{R,i} \gamma^{\mu} u_{R,i} \bar N_R
\gamma_{\mu} l_{R,i} + \alpha_{V_1}^{(i)} \bar l_{R,i} \gamma^{\mu} l_{R,i} \bar N_R \gamma_{\mu} N_R + \alpha^{(i)}_{V_2} \bar
L_i \gamma^{\mu} L_i \bar N_R \gamma_{\mu} N_R + \right. \nonumber
\\ && \left. \alpha^{(i)}_{V_3} \bar u_{R,i} \gamma^{\mu}
u_{R,i} \bar N_R \gamma_{\mu} N_R + \alpha^{(i)}_{V_4} \bar d_{R,i} \gamma^{\mu} d_{R,i} \bar N_R \gamma_{\mu} N_R +
\alpha^{(i)}_{V_5} \bar Q_i \gamma^{\mu} Q_i \bar N_R \gamma_{\mu} N_R + \right. \nonumber
\\ && \left.
\alpha^{(i)}_{S_0}(\bar \nu_{L,i}N_R \bar e_{L,i}l_{R,i}-\bar e_{L,i}N_R \bar \nu_{L,i}l_{R,i}) + \alpha^{(i)}_{S_1}(\bar
u_{L,i}u_{R,i}\bar N \nu_{L,i}+\bar d_{L,i}u_{R,i} \bar N e_{L,i})
 + \right. \nonumber
\\ && \left.
\alpha^{(i)}_{S_2} (\bar \nu_{L,i}N_R \bar d_{L,i}d_{R,i}-\bar e_{L,i}N_R \bar u_{L,i}d_{R,i}) + \alpha^{(i)}_{S_3}(\bar
u_{L,i}N_R \bar e_{L,i}d_{R,i}-\bar d_{L,i}N_R \bar \nu_{L,i}d_{R,i}) + \right. \nonumber
\\ && \left.  \alpha^{(i)}_{S_4} (\bar N_R \nu_{L,i}~\bar l_{L,i} N_R~+\bar
N_R e_{L,i} \bar e_{L,i} N_R) + \cdots  + h.c. \right\}
\end{eqnarray}
}
{In Eqs. \eqref{leff_svb} and \eqref{leff_4-f} the index $i$ label family and a sum is understood. The generic constants
$\alpha_{\mathcal J}$ with $\mathcal{J} \equiv NN\phi , \; LN\phi , \; Ne\phi , \; duNe , \; etc$ are re-labeled in order to simplify the
writing of the equations:}
\begin{eqnarray}
\alpha_Z&=&\alpha_{NN\phi},\; \alpha_{\phi}=\alpha_{LN\phi},\; \alpha_W=\alpha_{Ne\phi},\;
\alpha_{V_0}=\alpha_{duNe},\;\;
\alpha_{V_1}=\alpha_{eNN},\;\nonumber \\
\alpha_{V_2}&=&\alpha_{LNN},\;\alpha_{V_3}=\alpha_{uNN},\;
\alpha_{V_4}=\alpha_{dNN},\;\alpha_{V_5}=\alpha_{QNN},\;
\alpha_{S_0}=\alpha_{LNe},\;\nonumber \\
\alpha_{S_1}&=&\alpha_{QuNL},\; \alpha_{S_2}=\alpha_{LNQd},\;\;
\alpha_{S_3}=\alpha_{QNLd},\; \alpha_{S_4}=\alpha_{LN}.
\end{eqnarray}
{ where for simplicity in the notation we omit here the family index.
The one-loop generated operators are suppressed by the $1/(16 \pi^2)$ factor but, as we show in \cite{Duarte:2015iba}, these play a major role in the
$N$-decay. In particular for the low $m_N$ range studied here and for the relative size of the coupling constants we will discus in the section \ref{resultados}, the dominant channel decay is  $N \rightarrow \nu \gamma$, which is produced by different terms coming from the operators in \eqref{eq:ope3}.}
\begin{eqnarray}\label{leff_1loop}
\mathcal{L}_{eff}^{1-loop}&=&\frac{\alpha^{(i)}_{L_1}}{\Lambda^2} \left(-i\sqrt{2} v c_W P^{(A)}_{\mu} ~\bar \nu_{L,i} \sigma^{\mu\nu} N_R~ A_{\nu} 
+i \sqrt{2} v s_W P^{(Z)}_{\mu} ~\bar \nu_{L,i} \sigma^{\mu\nu} N_R~ Z_{\nu}+  \right)  
\nonumber 
\\ &-& \frac{\alpha^{(i)}_{L_2}}{\Lambda^2} \left(\frac{m_Z}{\sqrt{2}}P^{(N)}_{\mu} ~\bar \nu_{L,i} N_R~ Z^{\mu}+ 
+ m_W P^{(N)}_{\mu} ~\bar l_{L,i} N_R~ W^{-\mu} \right) 
\nonumber 
\\ &-& \frac{\alpha^{(i)}_{L_3}}{\Lambda^2}\left(i\sqrt{2} v  c_W P^{(Z)}_{\mu} ~\bar \nu_{L,i} \sigma^{\mu\nu}N_R~ Z_{\nu} 
+ i\sqrt{2} v s_W P^{(A)}_{\mu} ~\bar \nu_{L,i} \sigma^{\mu\nu}N_R~ A_{\nu} \right.
\nonumber 
\\ &+& \left. i 2\sqrt{2} m_W ~\bar \nu_{L,i} \sigma^{\mu\nu} N_R~ W^+_{\mu}W^-_{\nu} + i \sqrt{2} v P^{(W)}_{\mu} ~\bar l_{L,i} \sigma^{\mu\nu} N_R~ W^-_{\nu} \right.
\nonumber
\\ &+& \left. i 4 m_W c_W ~\bar l_{L,i} \sigma^{\mu\nu} N_R~ W^-_{\mu} Z_{\nu}+ i 4 m_W s_W ~\bar l_{L,i} \sigma^{\mu\nu} N_R~ W^-_{\mu} A_{\nu} 
\right)
\nonumber
\\ &-&  \frac{\alpha^{(i)}_{L_4}}{\Lambda^2} \left( \frac{m_Z}{\sqrt{2}} P^{(\bar\nu)}_{\mu}~\bar \nu_{L,i} N_R~ Z_{\mu}- 
\frac{\sqrt{2} m^2_W}{v} ~\bar \nu_{L,i} N_R~  W^{-\mu}W^+_{\mu} 
-\frac{m^2_z}{\sqrt{2} v} ~\bar \nu_{L,i} N_R~ Z_{\mu}Z^{\mu} \right.
\nonumber
\\ &+& \left.  m_W P^{(\bar l)}_{\mu} W^{-\mu} ~\bar l_{L,i} N_R + 
 e m_W  ~\bar l_{L,i} N_R W^{-\mu}A_{\mu} + e m_Z s_W ~\bar l_{L,i} N_R W^{-\mu}Z_{\mu} 
\right) + h.c.
\end{eqnarray}

where $P^{(a)}$ is the 4-moment of the incoming $a$-particle. Moreover $c_W=\cos(\theta_W)$ and $s_W=\sin(\theta_W)$ with
$\theta_W$ the Weinberg angle.

The constants $\alpha^{(i)}_{L_j}$, with $j=1,4$, are associated to the specific operators:
{
\begin{eqnarray}
\alpha^{(i)}_{L_1}=\alpha^{(i)}_{NB},\;\; \alpha^{(i)}_{L_2}=\alpha^{(i)}_{DN},\;\; \alpha^{(i)}_{L_3}=\alpha^{(i)}_{NW},\;\; 
\alpha^{(i)}_{L_4}=\alpha_{\bar DN}.
\end{eqnarray} 
}
The complete Lagrangian for the effective model is presented in an appendix in the recent work \cite{Duarte:2016miz}.
{
For completeness we include in Fig.\ref{fig:branching} a plot with the principal decay channel in the low mass region. A sum on particle
and antiparticle is understood and the channel $N \rightarrow \nu \, lep$ include the three body decays to leptons $N \rightarrow \nu_{l_i} l_i^+ l_i^-, \; \nu_{l_i} l_j^+ l_i^-
, \; \nu_{l_i} \nu_{l_i} \bar \nu_{l_i}$ 
}

In Sec. \ref{resultados}, we discuss the different bounds we consider on the coupling $\alpha^{(i)}_{\mathcal{J}}$ and the strategy we follow to take it into account 
for the prediction on the fluxes.

\section{Photon flux by heavy neutrino decay in the atmosphere.\label{photonproduction}}

We consider the transport equation for charged pions and kaons in the atmosphere including the interaction and decay terms \cite{Gaisser:2016uoy}:

\begin{equation}
\frac{d\phi_M(E,X)}{dX}=-\phi_M(E,X) \left(\frac1{\Lambda_M}+\frac{c_M}{E \rho(X)}\right)+\frac{Z_{\mathcal{N}M}}
{\lambda_{\mathcal{N}}}N_0(E) e^{(-X/\Lambda_{\mathcal{N}})},
\end{equation}
where $X$ is the slant path in ${\rm g/cm^2}$,  
 $\rho(X)$ is the density of the atmosphere, $M$ represents the meson $\pi$ or $K$ and $\mathcal{N}$ a nucleon. The cosmic nucleon flux is parametrized as
\begin{equation}
N_0(E)=A_0 E^{-(\gamma+1)}
\end{equation}
{ where $\gamma=1.7$ and $A_0=1.8$ is the normalization constant for the initial cosmic flux. The constants $c_{\pi}$ and $c_K$ are $m_{\pi}/\tau_{\pi}$ and $m_K/\tau_K$ respectively, with $\tau_{\pi,K}$ the mean lifetime of pion and kaon. 

The constants $Z_{i \; j}$ are the spectrum-weighted moments for the inclusive cross section for a incident particle $i$ colling with air nucleus and producing a 
outgoing particle $j$, with $i,\;j=\pi , K , \mathcal{N}$. On the other hand the attenuation length  constants $\Lambda_{\mathcal{N}}$, $\Lambda_{\pi}$, $\Lambda_K$ are related to the interaction length $\lambda_i$ by $\Lambda_i=\lambda_i (1-Z_{i\;i})$, with $i=\mathcal{N}, \; \pi, \; K$ \cite{Gaisser:2016uoy}}. 

With the usual approximations: (i) the hadron flux can be factorized $\phi_{\mathcal{N}}(E,X)=E^{-\alpha}\phi_{\mathcal{N}}(X)$, (ii) the interaction length is independent of energy and (iii) the differential cross section is Feynman scaling, the solution for the meson flux is \cite{Gaisser:2016uoy}
\begin{equation} \label{mesonflux}
\phi_M(E,X)=\left(\frac{A_0 Z_{\mathcal{N} M}}{\lambda_{\mathcal{N}}}\right) E^{-(\gamma+1)} X \int_0^1 du\; exp\left(-\frac{(1-u)X}{\Lambda_M}-\frac{X u}{\Lambda_{\mathcal{N}}}
-\frac{c_M}{E} k(X u,X)\right),
\end{equation}

where
\begin{equation}
k(X^{\prime},X)=\int_{X^{\prime}}^X \frac{dX^{\prime\prime}}{\rho(X^{\prime\prime})}.
\end{equation}
The transport equation for the heavy neutrino $N$ has as dominant contributions the absorption due to the $N$-decay and $N$-regeneration coming from the decay of mesons $\pi^{\pm}$ and $K^{\pm}$. We consider only the dominant meson decay process $M \rightarrow N \mu$, where $M$ represents the meson $\pi$ or $K$. The calculation of these decays in the effective theory we consider is shown in the Appendix \ref{ap1} and the obtained result for the width is:
\begin{eqnarray}
\Gamma^{M\rightarrow \mu N}&=&\frac{1}{16\pi m_M}\left( \frac{V^{uq} f_M m_M^2}{2 \Lambda^2} \right)^2 \left\{(\alpha_w^2+\alpha_{V_0})^2 \left[
(1+B_{\mu}-B_N)(1-B_{\mu}+B_N) \right. \right.
\nonumber \\
&-& \left.  (1-B_{\mu}-B_N)\right] + (\alpha_{S_2}+\alpha_{S_3}/2)^2 \frac{(1-B_{\mu}-B_N)}{(\sqrt{B_u}+\sqrt{B_q})^2} 
\nonumber \\
&+& \left. 2(\alpha_w+\alpha_{V_0})(\alpha_{S_2}+\alpha_{S_3}/2) \frac{\sqrt{B_{\mu}}(1-B_{\mu}+B_N)}{(\sqrt{B_u}+\sqrt{B_q})}    \right\}
\nonumber \\
& \times & \sqrt{(1-B_{\mu}+B_N)^2-4 B_N}
\end{eqnarray}
where
\begin{eqnarray}
B_{\mu} &=& m_{\mu}^2/m_M^2 \;\; , \;\; B_{N}=m_{N}^2/m_M^2 \;\; ,
\nonumber \\
B_{u} &=& m_{u}^2/m_M^2 \;\; , \;\; B_{q}=m_{q}^2/m_M^2 \;\; ,
\end{eqnarray}
being $q=d,s$ for $\pi$ or $K$, respectively.


Continuing with the transport equations, we have for the Majorana neutrino $N$

\begin{eqnarray}\label{transportN}
\frac{d\phi_N(E,X)}{dX}&=&-\frac1{\lambda^N_{dec}(E,X)}\phi_N(E,X)+Br(\pi\rightarrow N\mu)
\int_{z^{\pi}_{min}}^{z^{\pi}_{max}} \frac{dz}{z}\frac{\phi_{\pi}(E/z,X)}{\lambda^{\pi}_{dec}(E/z,X)}\frac{dn^{\pi}}{dz}+
\nonumber \\ 
&& Br(K\rightarrow N\mu) \int_{z^{K}_{min}}^{z^{K}_{max}} \frac{dz}{z}\frac{\phi_{K}(E/z,X)}{\lambda^{K}_{dec}(E/z,X)}\frac{dn^{K}}{dz},
\end{eqnarray}
with an absorption term given by the $N$-decay and two source terms coming from the meson decay, and where
\begin{eqnarray}
z^{\pi}_{min}=\frac12 (1+P_N-P_{\mu}-\sqrt{(1+P_N-P_{\mu})^2-4 P_N}) \nonumber \\
z^{\pi}_{max}=\frac12 (1+P_N-P_{\mu}+\sqrt{(1+P_N-P_{\mu})^2-4 P_N}) \nonumber \\
z^{K}_{min}=\frac12 (1+K_N-K_{\mu}-\sqrt{(1+K_N-K_{\mu})^2-4 K_N}) \nonumber \\
z^{K}_{max}=\frac12 (1+K_N-K_{\mu}+\sqrt{(1+K_N-K_{\mu})^2-4 K_N}) ,
\end{eqnarray}
with
\begin{equation}
P_i=(m_i/m_{\pi})^2 \;\; , \;\; K_i=(m_i/m_{K})^2.
\end{equation}
The branching ratios $B_r(\pi \rightarrow N\mu)$ and $B_r(K \rightarrow N\mu)$ can be written as 
$B_r(\pi \rightarrow N\mu)=( \Gamma(\pi \rightarrow N\mu)/\Gamma(\pi \rightarrow \nu\mu) ) \times B_r(\pi \rightarrow \nu\mu)$ and
$B_r(K \rightarrow N\mu)=( \Gamma(K \rightarrow N\mu)/\Gamma(K \rightarrow \nu\mu) ) \times B_r(K \rightarrow \nu\mu)$, with
$B_r(\pi \rightarrow \nu\mu)\sim 1$ and $B_r(K \rightarrow \nu\mu)\sim 0.64$.

Finally, the expressions for the decay distributions are
\begin{eqnarray}
\frac{dn^{\pi}}{dz}=\frac1{\sqrt{(1+P_N-P_{\mu})^2-4 P_N}} \;\; , \;\; \frac{dn^{K}}{dz}=\frac1{\sqrt{(1+K_N-K_{\mu})^2-4 K_N}}, 
\end{eqnarray}
and 
\begin{eqnarray}
\lambda^N_{dec}(E,x)&=&\frac{E}{m_N}\tau_N \rho(x)\equiv \frac{E \rho(x)}{c_N}, 
\nonumber \\ 
\lambda^{\pi}_{dec}(E/z,x)&=&\frac{E}{z m_{\pi}}\tau_{\pi} \rho(x)
 \equiv \frac{E \rho(x)}{z c_{\pi}}
\\
\lambda^{K}_{dec}(E/z,x)&=&\frac{E}{z m_K}\tau_K \rho(x) \equiv \frac{E \rho(x)}{z c_{K}}
\nonumber
\end{eqnarray}

replacing in Eq.\ref{transportN} we have

\begin{eqnarray}\label{transportN1}
&&\frac{d\phi_N(E,X)}{dX}=-\frac{c_N}{E\rho(X)}\phi_N(E,X)+
 \nonumber \\ && \frac{c_{\pi}}{E \rho(X)}\frac{Br(\pi\rightarrow N\mu)}{\sqrt{(1+P_N-P_{\mu})^2-4 P_N}}
\int_{z^{\pi}_{min}}^{z^{\pi}_{max}} dz\phi_{\pi}(E/z,X)+
\\ 
&& \frac{c_{K}}{E \rho(X)}\frac{Br(K\rightarrow N\mu)}{\sqrt{(1+K_N-K_{\mu})^2-4 K_N}}
\int_{z^{\pi}_{min}}^{z^{\pi}_{max}} dz\phi_{K}(E/z,X)
\nonumber
\end{eqnarray}
and for the integrals on the meson flux $\int_{z^{\pi}_{min}}^{z^{\pi}_{max}} dz\phi_{\pi}(E/z,X)$ and $\int_{z^{\pi}_{min}}^{z^{\pi}_{max}} dz\phi_{K}(E/z,X)$ we insert
the corresponding meson flux obtained in Eq.\ref{mesonflux}
\begin{eqnarray}
\int_{z^{M}_{min}}^{z^{M}_{max}} \phi_{M}(E/z,X) \, dz &=& \frac{A_0 Z_{\mathcal{N}M}}{\lambda_{\mathcal{N}}} E^{-(\gamma+1)} 
X \int_0^1 du \; \left(\int_{z^{M}_{min}}^{z^{M}_{max}} dz \, z^{(\gamma+1)}\, exp(-\frac{c_M}{E}k(X u,X) z) \right) 
\nonumber \\
&& \times exp(-\frac{(1-u)X}{\Lambda_M}-\frac{u X}{\Lambda_{\mathcal{N}}}),
\end{eqnarray}

with $M$ labeling the meson $\pi$ and $K$. The integration in the z-variable is direct:
\begin{eqnarray}
\int_{z^{M}_{min}}^{z^{M}_{max}} dz \, z^{(\gamma+1)}\, exp(-\frac{c_M}{E}k(X u,X) z)=E^{(\gamma+1)} {\cal H}^M(E,u,X) ,
\end{eqnarray}
\begin{figure}
\begin{center}
\includegraphics[width=0.8\textwidth]{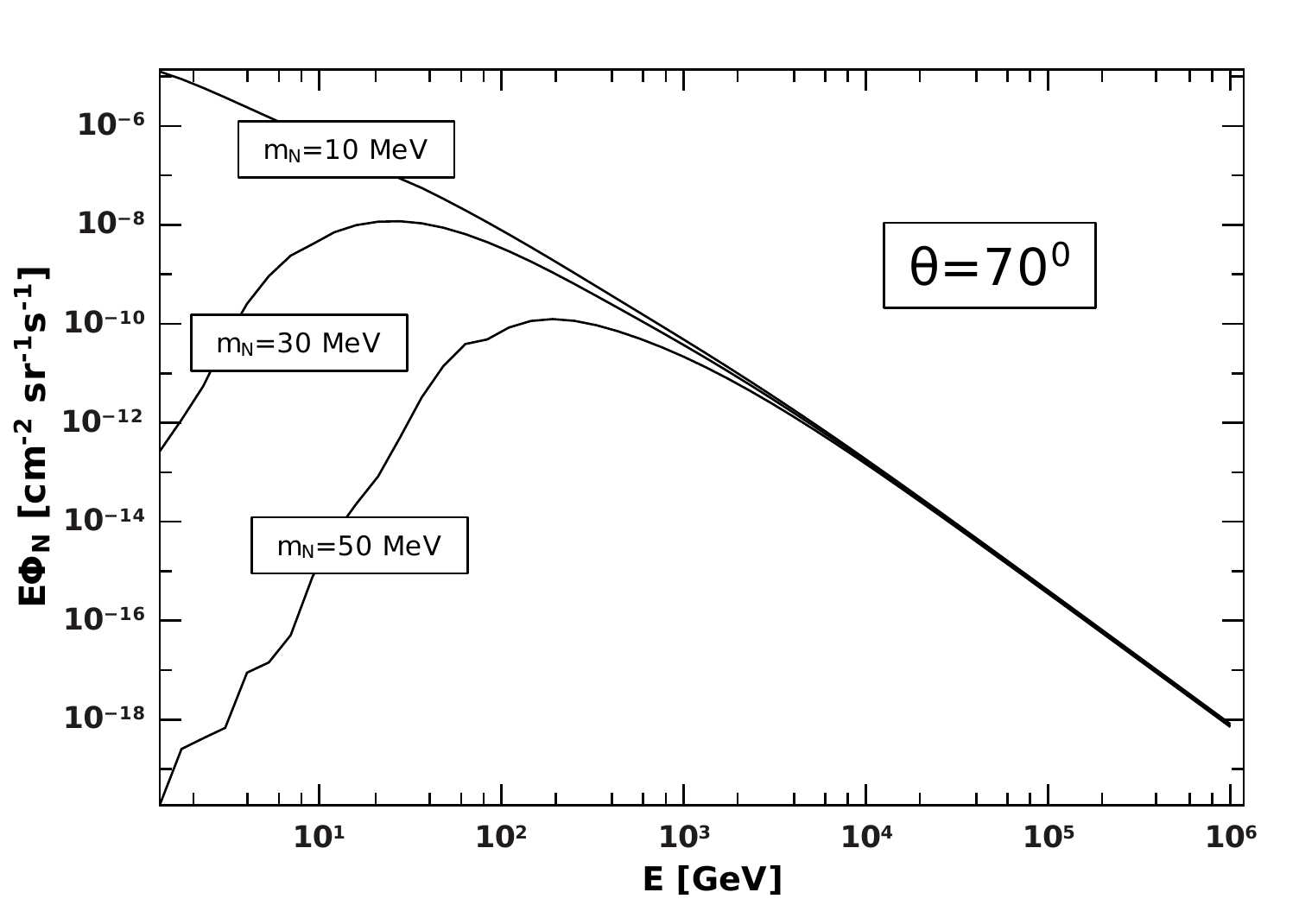}
\hspace{4cm}
 \caption{\label{fluxN} Heavy neutrino flux at the sea level as a function of the neutrino energy and for the intensity coupling indicated in the text.} 
\end{center}
\end{figure}
where the function ${\cal H}^M(E,u,X)$ reads
\begin{eqnarray}
{\cal H}^M(E,u,X)&=&\Gamma(\gamma+2)\left(c_M k(u X,X) \right)^{-(\gamma+2)} \nonumber \\
&& \times \left(\Gamma(\frac{c_M}{E}k(u X,X)z^M_{max},\gamma+2)-\Gamma(\frac{c_M}{E}k(u X,X)z^M_{min},\gamma+2) \right)
\end{eqnarray}
With the above definitions, we have the heavy neutrino flux given by
\begin{eqnarray}
\phi_N(E,X)&=&X^2 \int_0^1 \int_0^1 dv\, du\, \frac{v e^{-\frac{c_N}{E}k(v X,X)}}{\rho(v X)} \, e^{-\frac{u v X}{\Lambda_{\mathcal{N}}}}
\nonumber \\
&& \times \left[ {\cal D}_{\pi} {\cal H}^{\pi}(E,u,v X)\, e^{-\frac{(1-u)v X}{\Lambda_{\pi}}} 
+
{\cal D}_{K} {\cal H}^{K}(E,u,v X)\, e^{-\frac{(1-u)v X}{\Lambda_{K}}} \right],
\end{eqnarray}
where 
\begin{eqnarray}
{\cal D}_{\pi}=\dfrac{A_0 Z_{\mathcal{N}\pi} c_{\pi} B_r(\pi \rightarrow N\mu)}{\lambda_{\mathcal{N}}\sqrt{(1+P_N-P_{\mu})^2-4 P_N}}
\nonumber \\
{\cal D}_{K}=\dfrac{A_0 Z_{\mathcal{N}K} c_{K} B_r(K \rightarrow N\mu)}{\lambda_{\mathcal{N}}\sqrt{(1+K_N-K_{\mu})^2-4 K_N}}.
\end{eqnarray}

We show in Fig.(\ref{fluxN}) the flux of heavy neutrinos at the sea level as a function of the energy, { for a slant distance calculated for an angle $\theta=70^0$ with respect to the zenith direction and for the couplings intensity discussed in the next section}.
{ For high energy, the fluxes are independent of the value of $m_N$, while for lower energy heavy neutrinos present a lower flux due to a shorter decay time.}

As can be seen from Fig. \ref{fig:branching} (see also \cite{Duarte:2016miz}), for the masses of $N$ considered (tens of MeVs), 
the dominant decay channel is $N \rightarrow \gamma \nu$. This channel decay was calculated in \cite{Duarte:2016miz}
\begin{eqnarray}
\label{photon_neutrino}
\Gamma^{N\rightarrow \nu_i (\bar \nu_{i}) \gamma}=\frac1{2\pi}\left(\frac{v^2}{m_N}\right)\left(\frac{m_N}{\Lambda}\right)^4(\alpha_{L_1}^{(i)}c_W+\alpha_{L_3}^{(i)}s_W)^2.
\end{eqnarray}  
Thus, the total width for the low mass region is
\begin{equation}
\label{ancho_total}
\Gamma_t=\sum^3_{i=1}\left( \Gamma^{N\rightarrow \gamma \nu_i} + \Gamma^{N\rightarrow \gamma \bar\nu_i}\right)
\end{equation}
In order to study the production of photons by the heavy neutrino decays, we consider the coupled transport equations
\begin{eqnarray}
\frac{d\phi_{\gamma}}{dl}(E,l)&=&\frac{c_N}{E} \int_0^1 dy \phi_N(\frac{E}{1-y},l) \frac{dn}{dy}
\nonumber \\
\frac{d\phi_N}{dl}(E,l)&=&-\frac{c_N}{E} \phi_N(E,l),
\end{eqnarray}
where $c_N=m_N/\tau_N$. The decay $N \rightarrow \nu \gamma$ (see Appendix \ref{ap2} for the calculation of $dn/dy$) gives the source for the photon flux (first equation) as well as a depletion in the $N$ flux (second equation). The mean lifetime of $N$ ($\tau_N$) is given by the inverse of the width in Eq.\ref{ancho_total},  which is the dominant channel.
Inserting the solution of the second equation,
\begin{equation}
\phi_N(E,l)=\phi^0_N(E) exp(-\frac{c_N}{E} l),
\end{equation}
into the first equation and solving, we have the solution
\begin{equation}
\phi_{\gamma}(E,l)=\phi^0_{\gamma}(E) \; + \; 2 \, \int_0^1 \; dy \; y \; \frac{(1-exp(-\frac{c_N}{E}l(1-y))}{1-y} \, \phi^0_N(\frac{E}{1-y}).
\end{equation}
Thus, if we call $l$ the distance between the obstacle and the detector and $l_i$ the traveled distance inside it, then we can write the photon flux 
arriving the detector as:
\begin{eqnarray}
\Delta\phi_{\gamma}&=&\phi_{\gamma}(E,l+l_i)-\phi_{\gamma}(E,l_i) 
\nonumber \\
&=& 2 \int_0^1 \;dy\;\frac{y}{1-y}\;exp(-\frac{c_N}{E}(1-y)l_i)
\left[1-exp(-\frac{c_N}{E}(1-y)l)\right]\; \phi_N^0(\frac{E}{1-y}),
\end{eqnarray}
where we have removed the photons produced inside the obstacle because they get absorbed. 
\begin{figure}
\begin{center}
\includegraphics[width=0.8\textwidth]{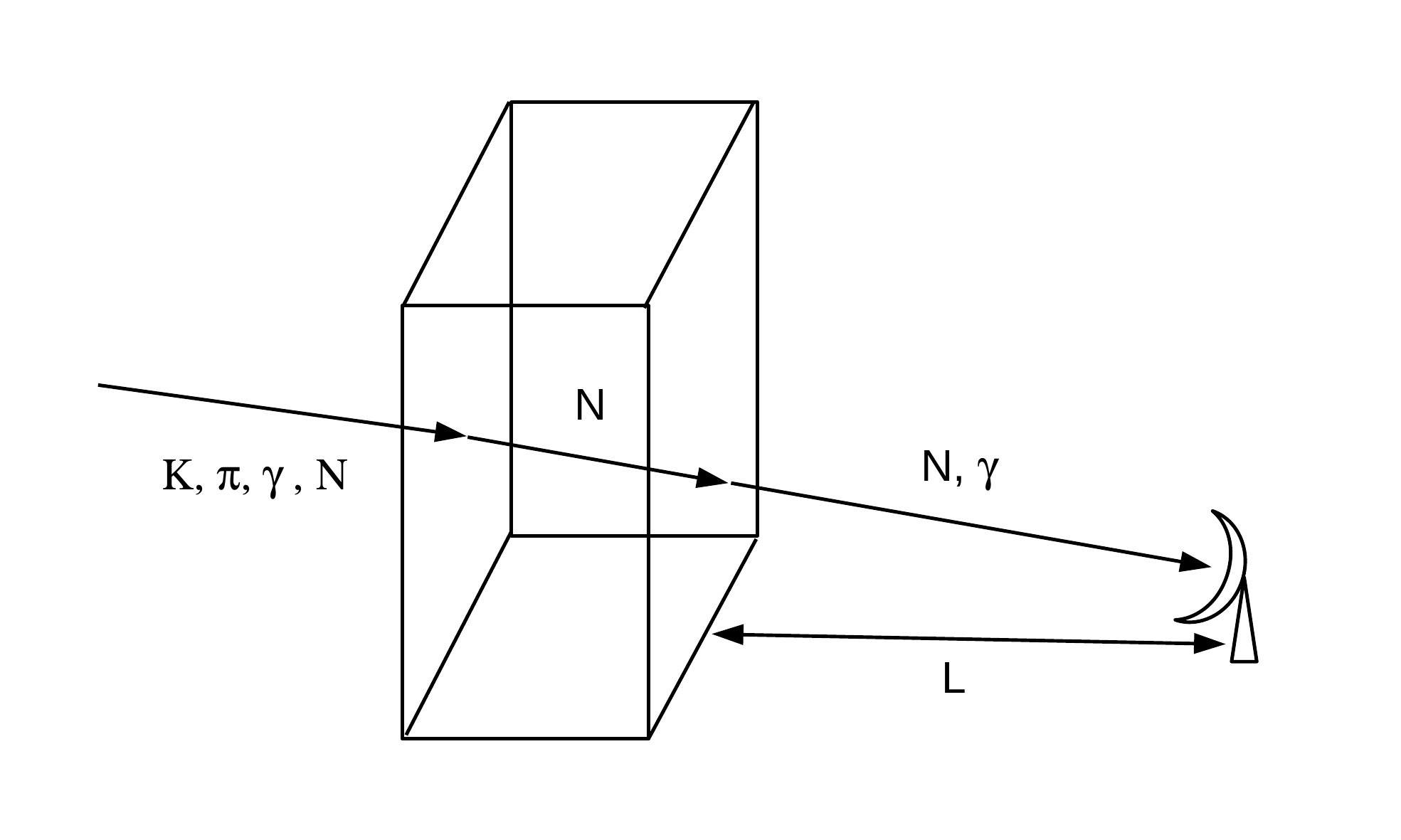}
\hspace{4cm}
 \caption{\label{fig:obstaculo} Schematic representation of the obstacle and the detector.} 
\end{center}
\end{figure}
In the next section we present our numerical results. We will consider the different bounds on the effective operators and the predictions for the photon flux, including the number of events to be detected.

\section{Numerical Results \label{resultados} }

{ The heavy Majorana neutrino couples to the three flavors families with couplings proportional to $\alpha^{(i)}_{\mathcal{J}}/\Lambda^2$. For
the case of the operator $\mathcal{O}^{(6),i}_{Ne\phi}$ these couplings can be related to the mixing angle between light and heavy neutrinos $U_{l_i N}$
in Eq.\ref{wlnu} \cite{delAguila:2008ir}
\begin{equation} \label{mixing}
U_{l_i N}=\frac{\alpha^{(i)}_W v^2}{2 \Lambda^2}
\end{equation}
In analogy we can use the combination $\alpha^{(i)}_{\mathcal{J}} v^2/(2 \Lambda^2)$ to represent the coupling intensity for all the operators.
As it was discussed \cite{Duarte:2014zea, Duarte:2015iba, Duarte:2016miz, Duarte:2016caz}, the most restrictive bound on the operators 
$\mathcal{O}^{(6),1}_{Ne\phi},\, \mathcal{O}^{(6),1}_{duNe}, \, \mathcal{O}^{(6),1}_{QuNL}, \, 
\mathcal{O}^{(6),1}_{LNQd},\, \mathcal{O}^{(6),1}_{QNLd}, \, \mathcal{O}^{(6),1}_{NW}$ 
involving the first family is placed by the $0\nu\beta\beta$-decay experimental result. With the definition in Eq.\ref{mixing} the bound on the mixing 
is translated on the coupling $\alpha^{(1)}_{\mathcal{J}}$ corresponding to the mentioned operators $\alpha^{(1)}_{\mathcal{J}} \leq \alpha^{bound}_{0\nu\beta\beta}= 3.2 \times 10^{-2} \left( m_N/(100 \ {\rm GeV}) \right)^{1/2}$
for $\Lambda=1$TeV. For the other operators, which are not included in the $0\nu\beta\beta$-decay, we consider for them the same bound corresponding to the $BELLE$ result \cite{Liventsev:2013zz}, $\alpha^{(i)}_{\mathcal{J}} \leq \alpha^{bound}_{BELLE} = 0.3$. Then, calling
\begin{equation}
U=\frac{\alpha^{(i)}_{\mathcal{J}} v^2}{2 \Lambda^2}
\end{equation}
we have $\vert U \vert^2\leq 8.8 \times 10^{-5}$ for the BELLE bound.

It is clear the different size between the contributions of both kind of operators. We maintain this hierarchy throughout the work decoupling the operators contributing
to $0\nu \beta\beta$. 

{ For the 1-loop generated operators we consider the coupling constants as $1/(16\pi^2)$ times the corresponding tree level coupling. Thus for the operators
$\mathcal{O}_{DW}$, $\mathcal{O}_{NW}$ and $\mathcal{O}_{\bar D N}$ which contribute to $0\nu\beta\beta$ we have
\begin{equation}
\alpha^{(1)}_{L_2}, \; \alpha^{(1)}_{L_3}, \; \alpha^{(1)}_{L_4}, \; \leq \frac{1}{16\pi^2}\alpha^{bound}_{0\nu\beta\beta}
\end{equation}  
for the first family.

For the operator $\mathcal{O}_{NB}$ which do not contribute to $0\nu\beta\beta$ we take 
\begin{equation}
\alpha^{(i)}_{L_1} \leq \frac1{16\pi^2}\alpha^{bound}_{BELLE}
\end{equation}

With these considerations and in order to estimate the intensity of the photon flux, we consider a generic  obstacle with thickness of $1$ km. We integrate the flux for the energy range $0<E<10^6$ GeV and consider the arriving direction as $\theta=70^0$ with respect to the zenith direction. We consider the received flux at different distances to the obstacle: 1 km, 5 km and 10 km. The results are shown in Fig.\ref{flu_gamma_dist_mN} as a function of $m_N$. 
\begin{figure}
\begin{center}
\includegraphics[width=0.8\textwidth]{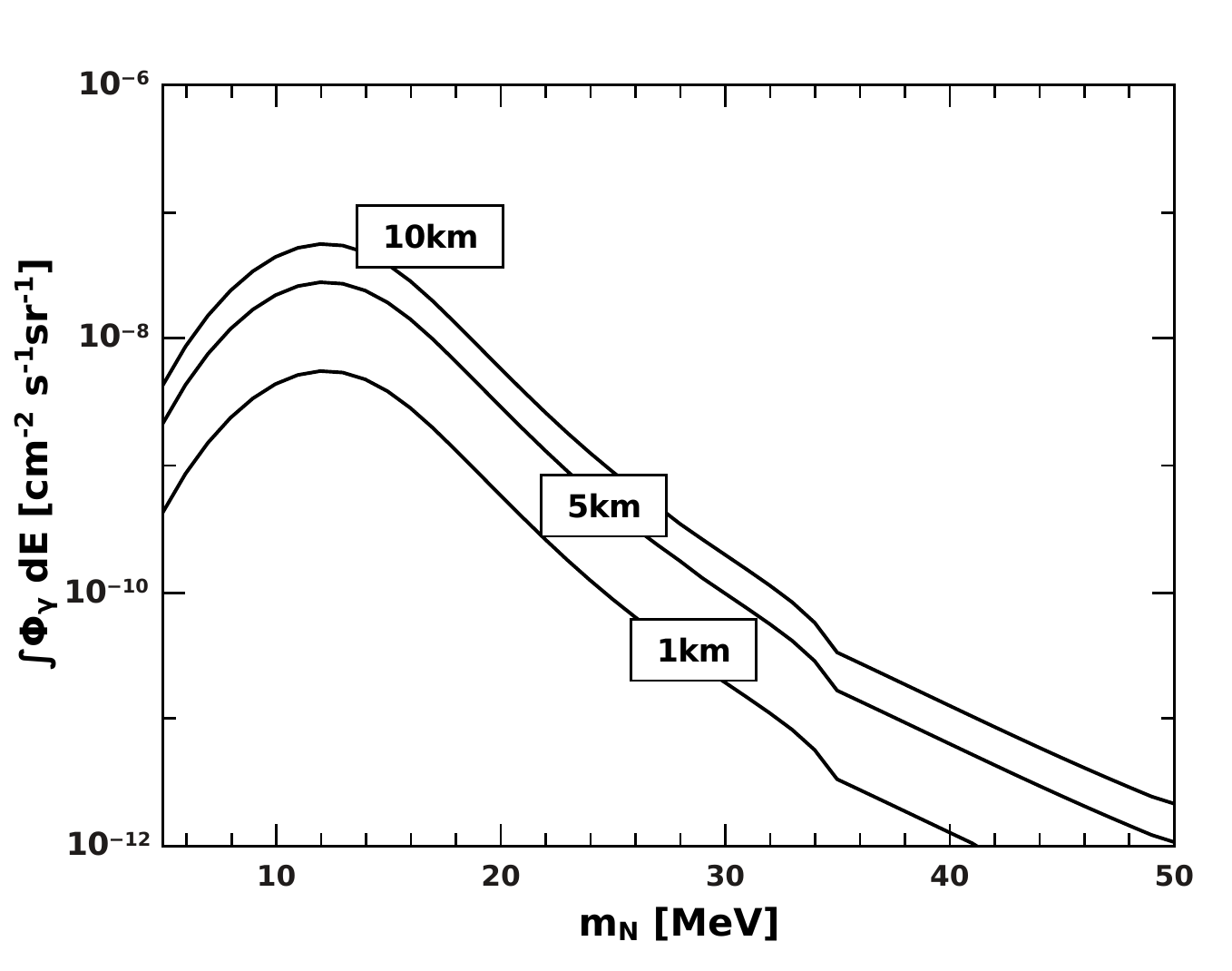}
\hspace{4cm}
 \caption{\label{flu_gamma_dist_mN} Photon flux integrated in energy as coming from a generic obstacle of $1$ km of thickness as a function of the N mass, and for different distances to the obstacle.} 
\end{center}
\end{figure}
Additionally, we show in Fig. \ref{flu_gamma_dist_E} the photon flux coming from an obstacle with sides of $1$ km (see Fig. \ref{fig:obstaculo} for a sketch) as a function of the energy for different neutrinos masses and for distances to the obstacle of 5 km and 10 km. We have integrated over the solid angle subtended by the obstacle.

\begin{figure}
\begin{center}
\includegraphics[width=1\textwidth]{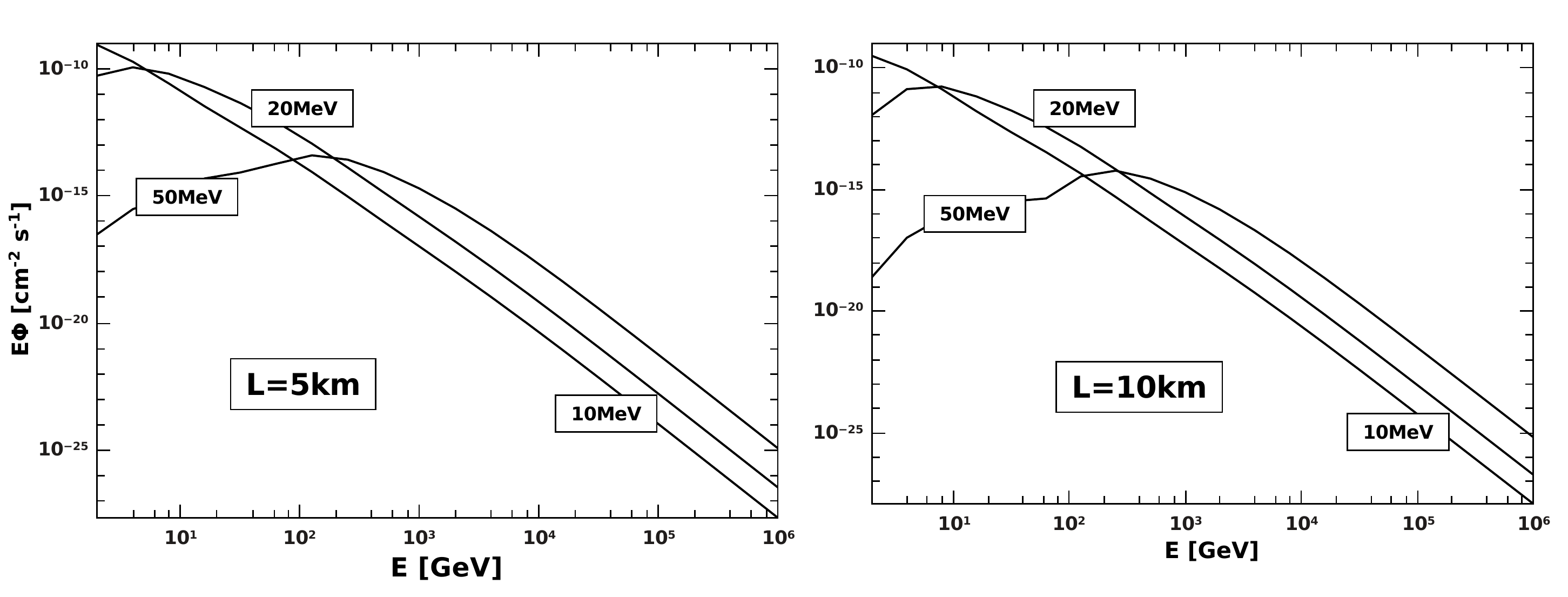}
\hspace{4cm}
 \caption{\label{flu_gamma_dist_E} Photon flux as coming from the obstacle described in the text as a function of the photon energy for different distances to the obstacle and different heavy neutrino masses.} 
\end{center}
\end{figure}
These curves were obtained for generic couplings compatible with $0\nu\beta\beta$ and the $BELLE$ bound. The idea here is to show the relative behavior of these fluxes for different masses
and distances to the obstacle. As we will discuss, there are most restrictive bounds that we will take into account when we show the final flux of photons.

In the considered mass range of tens of MeVs, the main sources of experimental bounds on the effective coupling $\alpha^{(i)}_{\mathcal{J}}$ are the
pion decay \cite{deGouvea:2015euy}, the beam dump experiments \cite{Drewes:2015iva}, astrophysical observations, and the non-observation in Super-Kamiokande 
\cite{Nakayama:2003ad,Kusenko:2004qc} of an excess of events coming from the decay of heavy neutrinos produced in the atmosphere. }

Very stringent bounds on the interaction of heavy neutrinos in the MeV mass range were obtained from primordial nucleosynthesis \cite{Drewes:2015iva}. These limits are typically valid under the assumption that $N$ is a relatively long-lived particle ($\tau_N > 0.01$ s) and with the dominant decay mode $N \rightarrow \nu e^+ e^-$, i.e., into an active neutrino and a $e^+ e^-$ pair.   These two conditions are not satisfied in our cases, where the lifetime is shorter in a large part of the 
parameter space, and on the other hand, the dominant channel by several orders of magnitude is $N \rightarrow \nu \gamma$ in the mass range considered.
 In Figs. \ref{fig:eve_5km} and \ref{fig:eve_10km},
we show the curve where the lifetime is $\tau_N=10^{-2}$ s, which is the limit required by cosmic and astrophysics bounds. For reference we also include the curve for $\tau_N=10^{-8}$ s.
On the other hand, the clear dominance of the neutrino plus photon channel makes the beam dump result inapplicable, as this decay mode to invisible particles is not considered in those analysis, and can considerably alter the number of events found for $N$ decays inside the detector
\cite{Deppisch:2015qwa,Drewes:2015iva}.

Another independent constraint on the effective operator coupling can be set based on the non-observation of heavy neutrino decays by the Super-K experiment. The heavy neutrinos produced by meson decays into the atmosphere would generate an excess of events in the detector. 
In order to estimate the importance of such effect, we calculate the fraction of neutrinos $N$ that arrive at the Earth's surface and could decay inside the detector, which is located one kilometer deep ($L_{deep}$) and has a forty meters long edge ($L_{e}$). The flight distance is a function of the
coupling and the $N$ mass, $L_{decay}=L_{decay}(m_N,U^2)$. We calculate the fraction $\eta(m_N,U^2)$ as the ratio between the number of heavy neutrinos decaying inside the detector and
the number of heavy neutrinos arriving at the Earth surface: 
\begin{eqnarray}
\eta(m_N,U^2)=\dfrac{\mathcal{N}^{(N)}_{detector}}{\mathcal{N}^{(N)}_{surface}},
\end{eqnarray}
where
\begin{eqnarray}
\mathcal{N}^{(N)}_{surface}=\idotsint \Phi_N^{sup}(E,\theta)\,dE\,da\,d\Omega\,dt
\end{eqnarray}
and
\begin{eqnarray}
\mathcal{N}^{(N)}_{detector}=
\idotsint \Phi_N^{sup}(E,\theta) \, exp\left(-\frac{L_{deep}m_N}{L_{decay} E}\right)\left[1- exp\left(-\frac{L_e m_N}{L_{decay} E}\right)\right] dE\,da\,d\Omega \,dt.
\nonumber \\
\end{eqnarray}
%

We have considered the data reported in \cite{Nakayama:2003ad} and the discussion in \cite{Kusenko:2004qc}. We have found that a factor $\eta=10^{-3}$ is a conservative value to impose the expected decay rates inside the detector does not exceed the rate of events detected by Super-Kamiokande \cite{Nakayama:2003ad}  experiment.

In the plots of Figs.\ref{fig:eve_5km} and \ref{fig:eve_10km}, we show the curve for which $\eta=10^{-3}$, which is a strong suppression factor. As we will see shortly there are regions of the parameters space where we still have an appreciable number of events and
less one in a thousand heavy neutrinos $N$ arriving the Earth decay inside the detector.
In the same figure we include the upper limit for the coupling as obtained from the pion decay \cite{deGouvea:2015euy}.

One further comment is in order at this point. In the Appendix \ref{ap1}, we show the expression for the meson decay in the context of the effective theory we are studying. In this expression, we can see a strong contribution from scalar operators due to the light quarks masses in the corresponding denominators.
In order to simplify the discussion, we will consider all the constants $\alpha_{\mathcal{J}}$ to be equal, but we have to take into account this important factor that determines the relative importance 
between the scalar and vectorial operators. 
For the pion decay, the coupling of scalar operators
 $\alpha_{scalar} \rightarrow \alpha_{S_2}, \alpha_{S_3}$ are accompanied by the big mass ratio $\frac{m_{\pi}}{(m_u+m_d)}$. 
Thus, is convenient to use the combination $(m_{\pi}/(m_u+m_d)) \times \alpha_{scalar}$  to compare with the experimental bound. If we call $\alpha_{bound}$ the corresponding bound, the value for $\alpha_{scalar}$ to use in the production of $N$ by meson decay is $\alpha_{scalar}=((m_u+m_d)/m_{\pi}) \times \alpha_{bound}$. 

In the case of $N$ production by pion decay, we replace $( m_{\pi}/(m_u+m_d) )\times \alpha_{scalar} \rightarrow \alpha_{bound}$. But in the case of $N$ production by $K$-decay, the replacement is $(m_K/m_s) \times \alpha_{scalar} \rightarrow (m_K/m_s) \times ((m_u+m_d)/m_{\pi}) \times \alpha_{bound} = 0.26 \times \alpha_{bound}$.

We present the results for number of events that could be detected by a Cherenkov telescope through the observation of the electromagnetic showers originated by $N$ decays after they have traversed an obstacle. In Figs.(\ref{fig:eve_5km}) and (\ref{fig:eve_10km}), we show the results as a contour plot for different number of events in the plane ($m_N , U2$) for different distances to the obstacle. 
We consider a generic detector like SHALON \cite{Sinitsyna:2013hmn} with an effective area of $10$ m$^2$, the solid angle spanned by the obstacle (which is compatible with the field of view of SHALON), and a detection time of one year. It is clear that, from the results shown, it is easy to obtain the number of events for different values of the observation time, effective area and number of detectors.
\begin{eqnarray}
\mathcal{N}^{(\gamma)}=\idotsint \Delta\phi_{\gamma}(E,\theta)\,dE\,da\,d\Omega\,dt
\end{eqnarray}
In the same figures, we show the bounds coming from pion decay and BELLE experiment \cite{Liventsev:2013zz}. Moreover we include the curves for which the $N$ lifetime is $10^{-2}$ s and for reference $10^{-8}$ s, as well as the curve for which 
$\eta=10^{-3}$. The arrows point to the allowed regions in the parameter space.
Our results are indicated by contour curves that correspond to different values of the number of events $1$, $10$ , $100$ and $1000$.
We can see that there are regions with an appreciable photon number of events satisfying experimental and observational bounds. The distances considered are enough for the development of
an electromagnetic cascade.
%
%

%
%

%
\begin{figure*}[h]
\begin{center}
\subfloat[]{\label{fig:eve_5km}\includegraphics[width=0.5\textwidth]{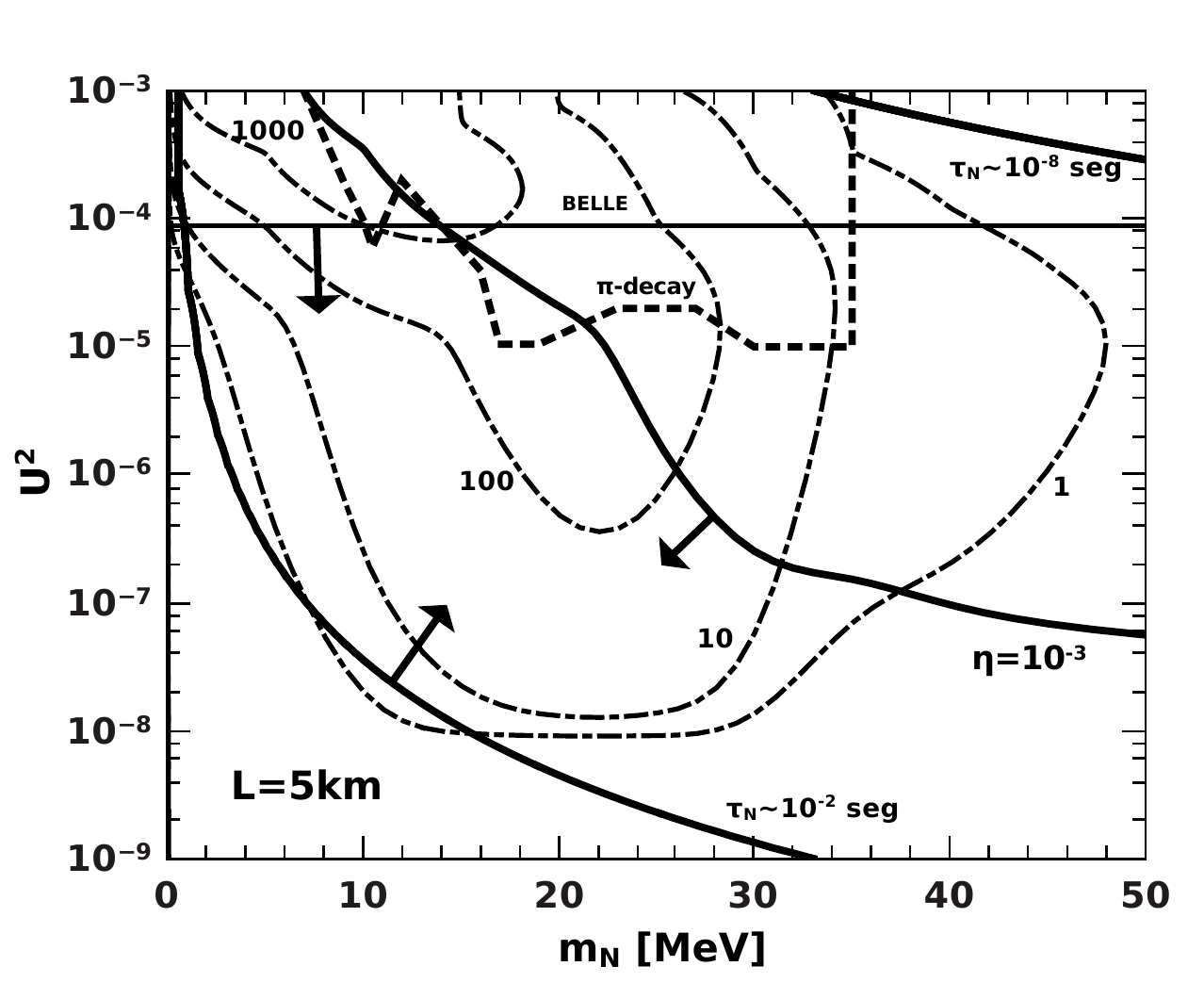}}
~
\subfloat[]{\label{fig:eve_10km}\includegraphics[width=0.5\textwidth]{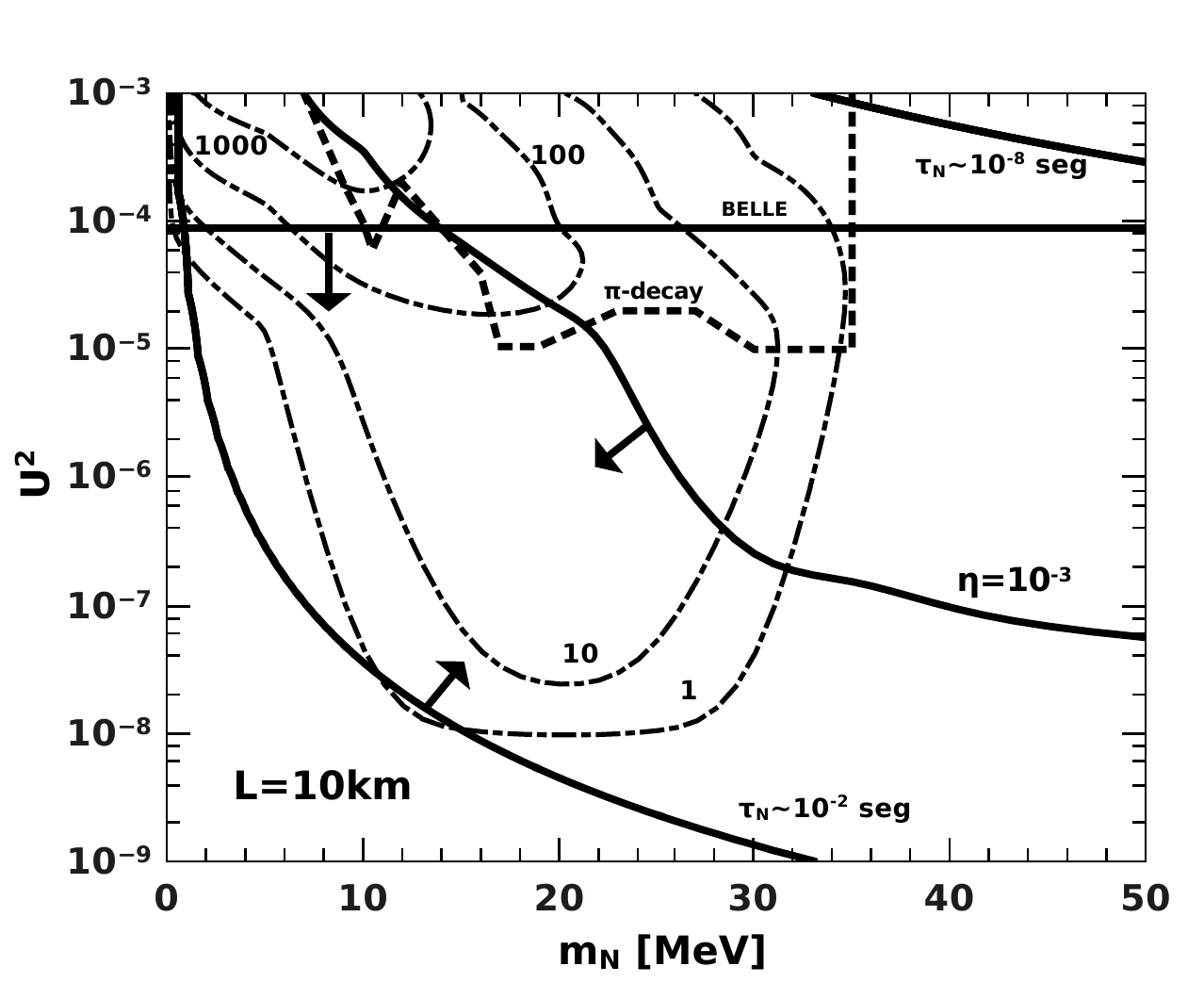}}
\caption{\label{fig:eventos} Contours for obtained number of events in the ($m_N,U^2$) plane are shown in dashed-dotted lines. We include bounds
coming form $\pi$-decay, cosmological bounds, these imposing by Super-K and BELLE experiment. The arrows indicate
the allowed region and the labels close to the dot-dashed curves the photon number of events.}
\end{center}
\end{figure*}

\section{Final remarks \label{conclusiones}}
We have studied the possibility to place reliable bounds to heavy neutrino couplings by considering their decay channel to photons once they have been produced by meson decays in the atmosphere. We have calculated the
photon flux originated by the decay of $N$'s emerging of an opaque object such as a mountain. These photons would be observable by a ground based Cherenkov instrument like SHALON \cite{Sinitsyna:2013hmn} or any other gamma-ray detector that could observe the showers initiated by such high energy photons coming in the direction of the obstacle. We considered different situations of obstacles placed at different distances to the detector. 
 The contours curves for the number of events are shown
in Figs. \ref{fig:eve_5km} and \ref{fig:eve_10km}, where we also include experimental and observational bounds. As it can be seen, there are regions in the $m_N-U^2$ space with a significant number of events and safe from experimental restrictions.

{\bf Acknowledgements} 

We thank CONICET and Universidad Nacional de Mar del
Plata (Argentina) for their 
financial supports.

\appendix

\section{Meson decay to Majorana neutrino in a effective theory.}\label{ap1}

In the context of the effective theory presented in Sec. \ref{modelo}, we calculate the contribution to the decay of mesons $\pi$ and $K$, generically $M$ decay. From the different Lagrangian presented in this section, we consider here the relevant pieces for the considered decay. Thus, we have
\begin{eqnarray}
\mathcal{L}&=&\frac{1}{\Lambda^2} \left\{-\alpha_w \frac{v m_w}{\sqrt{2}} \bar N_R \gamma^{\nu} \mu_R W^{\dagger}_{\nu}+
\alpha_{V_0} V^{ud} \bar d_R \gamma^{\nu} u_R \bar N_R \gamma_{\nu} \mu_R-\alpha_{S_2} V^{ud} \bar \mu_L N_R \bar u_L d_R  + \right.
\nonumber \\
&&\left. \alpha_{S_3} V^{ud} \bar u_L N_R \bar \mu_L d_R \; + \; hc \; + \; \cdots \; + \; d \; \rightarrow s \right\}.
\end{eqnarray}
We deduce the decay amplitude
\begin{eqnarray}
\mathcal{M}=-i \frac{V^{uq}}{\Lambda^2} 
&& \left\{ -\alpha_w \left<0\vert \bar u \gamma^{\nu} P_L q \vert M \right>  \left<N \mu \vert \bar \mu \gamma_{\nu} P_R N \vert 0\right> + 
\alpha_{V_0} \left<0\vert \bar u \gamma^{\nu} P_R q \vert M \right>  \left<N \mu \vert \bar \mu \gamma_{\nu} P_R N \vert 0\right> - \right.
\nonumber \\
&&\left.  \alpha_{S_2} \left<0\vert \bar u P_R q \vert M \right>  \left<N \mu \vert \bar \mu P_R N \vert 0\right> +
\alpha_{S_3} \left<N\mu\vert \bar u P_R N \bar \mu P_R q \vert M\right> \right\},
\end{eqnarray}
with $q=d, s$ for $\pi$ and $K$ decay, respectively.
In the last term, we need to rearrange the field operators in order to put together quarks fields in a sandwich and the lepton fields in another. 
In order to do that, we make a Fierz transformation to the last term taking into account a minus sign from the permutation of fermions, and then we have
\begin{eqnarray} \label{ampl}
\mathcal{M}=-i \frac{V^{uq}}{\Lambda^2} 
&& \left\{ -\alpha_w \left<0\vert \bar u \gamma^{\nu} P_L q \vert M \right>  \left<N \mu \vert \bar \mu \gamma_{\nu} P_R N \vert 0\right> + 
\alpha_{V_0} \left<0\vert \bar u \gamma^{\nu} P_R q \vert M \right>  \left<N \mu \vert \bar \mu \gamma_{\nu} P_R N \vert 0\right> - \right.
\nonumber \\
&&\left.  \alpha_{S_2} \left<0\vert \bar u P_R q \vert M \right>  \left<N \mu \vert \bar \mu P_R N \vert 0\right> -
\alpha_{S_3} \frac12 \left[ \left<0\vert \bar u P_R q \vert M \right>  \left<N \mu \vert \bar \mu P_R N \vert 0\right> + \right.\right.
\nonumber \\
&& \left. \left. \frac12 \left<0\vert \bar u \sigma^{\mu\nu}P_R q \vert M \right>  \left<N \mu \vert \bar \mu \sigma_{\mu\nu} P_R N \vert 0\right>\right]
\right\}
\end{eqnarray}
The calculation of the leptonic matrix element is straightforward, 
\begin{eqnarray}
\left<N \mu \vert \bar \mu \gamma_{\nu} P_R N \vert \; 0 \; \right> &=& \bar u_{\mu}(p_1) \gamma_{\nu} P_R v_N(p_N)
\nonumber \\
\left<N \mu \vert \bar \mu P_R N \vert \; 0 \; \right> &=& \bar u_{\mu}(p_1) P_R v_N(p_N)
\end{eqnarray}
In order to calculate the hadronic matrix element, we have to rely on the symmetries. The matrix element 
$\left<\;0\vert \bar u \gamma^{\nu} \gamma_5 q \vert \; M \; \right>$ is a Lorentz 4-vector because the meson $M$ is pseudoscalar and 
$\bar u \gamma^{\nu} \gamma_5 q$ is a pseudo 4-vector. 
The meson state is described by its four momentum $q^{\mu}$ and nothing else, since the pion has spin zero. Therefore, $q^{\mu}$ is the only 
4-moment on which the matrix element depends and it must be proportional to $q^{\mu}$. Thus, we can write
\begin{eqnarray}
\left<\;0\vert \bar u \gamma^{\nu} \gamma_5 q \vert \; M \; \right>  \;\; = \;\; i f_M q^{\nu}
\end{eqnarray} 
On the other hand, for the same reason, the matrix element of the 4-vector is
\begin{eqnarray}
\left<\;0\vert \bar u \gamma^{\nu} q \vert \; M \; \right>  \;\; = \;\; 0.
\end{eqnarray} 
In the case of the matrix element of the scalar or pseudo-scalar, we have to use the equation of motion
\begin{eqnarray}
\left<\;0\vert \bar u \gamma_5 q \vert \; M \; \right>  \;\; &=& \;\; -i \frac{m_M^2 f_M}{m_q+m_u}
\nonumber \\
\left<\;0\vert \bar u  q \vert \; M \; \right>  \;\; &=& \;\; 0
\nonumber \\
\left<\;0\vert \bar u \sigma_{\mu\nu} b \vert \; M \; \right>  \;\; &=& \;\; 0
\nonumber \\
\left<\;0\vert \bar u \sigma_{\mu\nu} \gamma_5 b \vert \; M \; \right>  \;\; &=& \;\; 0,
\end{eqnarray} 
where $m_q=m_d, m_s$ for the decay of $\pi$ and $K$, respectively.

Putting it all together and integrating over the 2-body phase space, we obtain 
\begin{eqnarray}
\Gamma^{M\rightarrow \mu N}=\frac{\vert \mathcal{M} \vert^2}{16\pi m_M^3}\sqrt{(m_M^2+m_N^2-m_{\mu}^2)^2-4m_M^2m_N^2},
\end{eqnarray}
with $\mathcal{M}$ given in Eq.\ref{ampl}.

The result is
\begin{eqnarray}
\Gamma^{M\rightarrow \mu N}&=&\frac{1}{16\pi m_M}\left( \frac{V^{uq} f_M m_M^2}{2 \Lambda^2} \right)^2 \left\{(\alpha_w^2+\alpha_{V_0})^2 \left[
(1+B_{\mu}-B_N)(1-B_{\mu}+B_N) \right. \right.
\nonumber \\
&-& \left.  (1-B_{\mu}-B_N)\right] + (\alpha_{S_2}+\alpha_{S_3}/2)^2 \frac{(1-B_{\mu}-B_N)}{(\sqrt{B_u}+\sqrt{B_q})^2} 
\nonumber \\
&+& \left. 2(\alpha_w+\alpha_{V_0})(\alpha_{S_2}+\alpha_{S_3}/2) \frac{\sqrt{B_{\mu}}(1-B_{\mu}+B_N)}{(\sqrt{B_u}+\sqrt{B_q})}    \right\}
\nonumber \\
& \times & \sqrt{(1-B_{\mu}+B_N)^2-4 B_N},
\end{eqnarray}
where
\begin{eqnarray}
B_{\mu} &=& m_{\mu}^2/m_M^2 \;\; , \;\; B_{N}=m_{N}^2/m_M^2 \;\; ,
\nonumber \\
B_{u} &=& m_{u}^2/m_M^2 \;\; , \;\; B_{q}=m_{q}^2/m_M^2 \;\; .
\end{eqnarray}

\section{$N \rightarrow \nu \gamma$ in the Laboratory}\label{ap2}

We follow the development shown for $\mu$-decay in the book of T. K.Gaisser \cite{Gaisser:1990vg}, but in our case for the $N\rightarrow \gamma \nu$ decay (we adapt the calculations presented in the Appendix of the recent work \cite{Duarte:2016smd}). 
First, we obtain the $N$ decay width in its rest frame, and then boost the result to the Laboratory frame. In the $N$ rest frame, we have the following expression: 
\begin{eqnarray}
\frac{1}{\Gamma_{\rm rest}}\frac{d\Gamma_{\rm rest}}{dx \, d\cos\theta_{\nu}}=  2\left(F_0(x)- P F_1(x)\cos\theta_{\nu}
\right),
\end{eqnarray}
being $\theta_\nu$ the direction of motion of the final $\nu$ taken from the Majorana neutrino $N$ moving direction, and
$P=\cos\theta_P$ where $\theta_P$ is the angle between the Majorana neutrino spin direction in its rest frame, and its moving direction as seen from the Laboratory frame. The variable $x$ represents the quotient between the final neutrino energy in the rest frame of the $N$ and the mass of the Majorana
neutrino: $x=k^0/m_N$.
The functions $F_0(x)$ and $F_1(x)$ are
\begin{eqnarray}
F_0(x)= x (1-x)\delta(x-1/2) \nonumber \\ \nonumber
\\
F_1(x)=x^2 \delta(x-1/2).
\end{eqnarray}
To obtain the corresponding expression in the laboratory frame, we make the appropriate Lorentz transformations. 
Denoting by $E_\nu$ and $E_N$ the Laboratory energies of the final neutrino and the Majorana neutrino, respectively, we have
\begin{equation}
z=x(1-\beta_{N} \cos\theta_\nu),
\end{equation}
with $z=E_\nu/E_N$ and $\beta_{N}=\sqrt{1-{m_N}^2/E^2_{N}}\simeq 1$.

We implement the Lorentz transformation with the help of the $\delta$-function, yielding
\begin{eqnarray}
\frac{1}{\Gamma_{\rm LAB}} \frac{d\Gamma_{\rm LAB}}{dz \, dx  \, d\cos\theta_\nu} =  2 \left(F_0(x)-P F_1(x)
\cos\theta_\nu \right) \delta\left[z-x\left(1+\beta_N \cos\theta_{\nu}\right)\right].
\end{eqnarray}
We first integrate over $\theta_\nu$ and then we integrate over $x$ in the interval $(x_{\rm min},x_{\rm max})$ with $x_{\rm
min}=z/(1+\beta_N)$ and $x_{\rm max}={\rm min}(1,z/(1-\beta_N))$, obtaining
\begin{eqnarray}
\frac{1}{\Gamma_{\rm LAB}}\frac{d\Gamma_{\rm LAB}}{dz}=2(1-z)\Theta(1/2-x(z)_{min})\Theta(x(z)_{max}-1/2).
\end{eqnarray}

For the low mass range considered in this work, the clearly dominant decay channel is the neutrino plus photon mode, and $\Gamma^{\rm
tot}_{LAB}(E)=\sum_{i=e,\mu,\tau}\left(\Gamma^{N\rightarrow \nu_{i} \gamma}_{LAB}(E)+\Gamma^{N\rightarrow \bar\nu_{i} \gamma}_{LAB}(E) \right)$. Then we consider the $\gamma$ decay channel, 
leading to the final $\gamma$ photon distribution in the laboratory frame:
\begin{eqnarray}
\frac{1}{\Gamma^{\rm tot}_{\rm LAB}(E)}\frac{d\Gamma^{N\rightarrow \nu (+\bar\nu)\gamma}_{\rm LAB}}{dz} \equiv\frac{dn(z)}{dz}.
\end{eqnarray}
Thus, after the indicated integrations in the evolution equations, the useful expression that we obtain is
\begin{eqnarray}
\frac{dn(z)}{dz}=\frac{n(1-y)}{dy}= 2 y,
\end{eqnarray}
where $z=1-y$, $x_0=1/2$ and $P=+1$ for the right-handed Majorana neutrinos.


\end{document}